\documentclass[fleqn,10pt]{wlscirep}
\usepackage[utf8]{inputenc}
\usepackage[T1]{fontenc}
\usepackage{etoolbox}
\usepackage{mathptmx}
\usepackage{bm}

\title{Computation of Biological Conductance with Liouville Quantum Master Equation}

\author[1,*]{Eszter Papp}
\author[1]{Gábor Vattay}
\affil[1]{Department of Physics of Complex Systems, Eötvös Loránd University, H-1053 Budapest, Egyetem tér 1-3.,Hungary}

\affil[*]{eszter.papp@ttk.elte.hu}

\begin{abstract}
Recent experiments have revealed that single proteins can display high conductivity, which stays finite for low temperatures, decays slowly with distance, and exhibits a rich spatial structure featuring highly conducting and strongly insulating domains. Here, we intruduce a new formula by combining the density matrix of the Liouville-Master Equation simulating quantum transport in nanoscale devices, and the phenomenological model of electronic conductance through molecules, that can account for the observed distance- and temperature dependence of conductance in proteins. We demonstrate its efficacy on experimentally highly conductive extracellular cytochrome nanowires, which are good candidates to illustrate our new approach by calculating and visualizing their electronic wiring, given the interest in the arrangement of their conducting and insulating parts.
As proteins and protein nanowires exhibit significant potential for diverse applications, including energy production and sensing, our computational technique can accelerate the design of nano-bioelectronic devices.
\end{abstract}
\begin{document}

\flushbottom
\maketitle

\thispagestyle{empty}

Protein electron transport measurements exhibit unique properties, making them excellent subjects for in-depth study and exploration.~\cite{amdursky2014electronic, bostick2018protein}. When electrodes are attached to protein structures, the measured conductance is surprisingly high, reaching nanoSiemens even over several nanometers of distance~\cite{zhang2017observation, zhang2019electronic, zhang2019role}. It is also noteworthy that the conductance remains stable even when the temperature changes from tens of Kelvins to ambient temperatures~\cite{kayser2019solid, sepunaru2011solid, bera2023near} and does not show significant decay with increasing protein size between the electrodes~\cite{zhang2019role, zhang2019electronic, bera2023near}. In bioelectronic measurements, where metallic contacts are chemically bound to molecules, molecular junctions are formed. The Landauer-Büttiker formula is a theoretical tool that accurately describes coherent elastic quantum transport, expressing the conductance in terms of the scattering matrix elements between metallic leads~\cite{lambert2015basic}. However, this formula is not applicable at high temperatures, and electron transfer is typically addressed through the semiclassical Marcus theory~\cite{marcus1956theory}. It is important to note that both theories are only limiting cases, and electron-vibrational (electron-phonon) interactions should be treated with care in the intermediate regime.

Here, we introduce a new, computationally accessible phenomenological approach that enables us to determine the conductance between atomic orbitals even in the intermediate temperature ranges and can pinpoint areas of high conductivity and insulation in any protein. We build on previous results~\cite{segal2000electron,davis1997electron}, where a quantum master equation has been used to calculate the electron transfer rate. The novelty is that using the approach developed in Refs.~\cite{zahid2003electrical,datta1997electronic} for the electric current in molecules, we derive a new formula connecting the master equation and the conductance. 
We then use the Liouville master equation introduced by Gebauer and Car~\cite{gebauer2004kinetic, fischetti1999master,gebauer2004current} for the reduced density matrix of electrons to describe electron transport in nanoscale systems. Its main advantage compared to other quantum master equations, such as the Lindblad and the Redfield equations~\cite{breuer2002theory}, is that in the absence of external perturbations, it drives the system toward the correct Fermi-Dirac distribution $F(E,\mu)=1/(1+e^{(E-\mu)/kT})$. This feature makes the Liouville master equation a more appropriate starting point for calculating conductance in
molecules. The result is a formula where the conductance is given in terms of the matrix elements of the single electron Hamiltonian, the couplings to the contacts, and the spectral density of the phonon bath. 

Following Refs.~\cite{davis1997electron,segal2000electron}, the reduced density matrix $\varrho_{nm}$ of an electron in a molecule can be described with the quantum master equation, which includes
the Hamiltonian $H_{nm}$ of the molecule in atomic orbital site basis, the escape rate $\Gamma_m/\hbar$ from the site $m$, 
the external current $J_m$ at the site $m$ and the operator $R_{nm}(\varrho)$, which is the phenomenological descriptor of the interaction with the phonon bath. This approach neglects cotunneling and is valid in the linear-response regime. The site-based quantum master equation can be transformed into the energy representation 
\begin{equation}
    \dot{\varrho}^{ij}=-\frac{i}{\hbar}(E_i-E_j)\varrho^{ij}- \frac{1}{2\hbar}\sum_r (\Gamma^{ir}\varrho^{rj}+\varrho^{ir}\Gamma^{rj})+R^{ij}(\varrho)+J^{ij},\label{energyrep}
\end{equation}   
with transformed matrix elements in the energy basis $\varrho^{ij}=\sum_{nm}\Psi_n^i\varrho_{nm}\Psi^j_m$,  $\Gamma^{ij}=\sum_n\Psi_n^i\Gamma_n\Psi_n^j$, $J^{ij}=\sum_n\Psi_n^iJ_n\Psi_n^j$ and $R^{ij}=\sum_{nm}\Psi_n^i\Psi_m^jR_{nm}(\varrho)$,
where $\Psi_n^i$ and $E_i$ are the eigenfunctions and the egienvalues $E_i\Psi_n^i=\sum_m H_{nm}\Psi_m^i$ of the Hamiltonian. Please, see
\textcolor{blue}{Supplementary Information S1} for details.
We assume that the Hamiltonian is an $N\times N$ real symmetric matrix with real eigenvalues and eigenvectors, where $N$ is the number of atomic orbitals. 
The operator $R(\varrho)$ describes the interaction with the environment and ensures the correct equilibrium properties of the electron 
distribution. In the framework of the single electron picture, the following Liouville master equation has been introduced~\cite{gebauer2004kinetic,gebauer2006kohn} to describe electron transport in nanoscale systems
\begin{equation}
    R^{ij}(\varrho)=(\delta_{ij}-\varrho^{ij})\sum_p \frac{1}{2\hbar}(\gamma^{ip}+\gamma^{jp})\varrho^{pp}
    -\varrho^{ij}\sum_p \frac{1}{2\hbar}(\gamma^{pi}+\gamma^{pj})(1-\varrho^{pp}),\label{Car}
\end{equation}
where $\gamma^{ij}$ are the transition rates between the energy levels. 
This equation can account for the exclusion principle, and in the absence of external currents and escape, its equilibrium solution is the Fermi-Dirac distribution
$\varrho^{ij}_{eq}=\delta_{ij}F(E_i,\mu)$, where $\mu$ is the chemical potential of the system. Transition rates between the electron
levels are
\begin{equation}
\gamma^{ij}=\gamma(\omega_{ij}) \sum_{n}|\Psi_n^i|^2|\Psi_n^j|^2,\label{transition}
\end{equation}
where $\omega_{ij}=(E_i-E_j)/\hbar$ is the transition frequency and $\gamma(\omega)$ is the spectral density of the phonon bath, that 
obeys the Boltzmann-type detailed balance equation $\gamma(\omega)/\gamma(-\omega)=e^{-\hbar\omega/kT}.$ The spectral density of proteins can be modeled effectively
by the Ohmic oscillator bath with cutoff\cite{mohseni2008environment}
$\gamma(\omega)=\eta\hbar\omega e^{-|\omega|/\omega_c}/(e^{\hbar\omega/kT}-1).$
The concrete form and parameters will not be used in this paper. For reference, the typical cutoff energy is $\hbar\omega_c\approx 0.0185$ $eV$, and the parameter $\eta=2\pi E_R/\hbar\omega_c\approx 1.46$ where $E_R$ is the reorganization energy\cite{mohseni2008environment}. Note that in order to maintain the validity of the Master equation under conditions of weak coupling between the electrons and phonons, the couplings $\gamma^{ij}$ should be small in comparison to the energies of the electrons and phonons. Conversely, the couplings $\Gamma^{ij}$ describe the electrons leaking from the molecule to the electrodes and are not subject to the same limitations.

\section*{Conductance}

One can couple contacts to two atomic orbital sites called left ($L$) and right ($R$) and calculate the conductance between them. For the derivation, please see the \textcolor{blue}{Supplementary Information S2}; here, we summarize only the main steps.
We switch on a small voltage difference $U$ between the contacts with escape rates $\Gamma_L/\hbar$ and $\Gamma_R/\hbar$, and their chemical potential shifts slightly to $\mu_{L/R}=\mu\pm eU/2$. According to the theory developed in Refs.\cite{zahid2003electrical,datta1997electronic}, the occupancy of energy $E_i$ in the left and right contacts changes to 
\begin{equation}
    \varrho_{eq}^i(\mu_{L/R})\approx\varrho_{eq}^i(\mu) \pm D^i(\mu)eU
\end{equation}
where $D^i(\mu)=\int dEf(E,\mu)d_i(E)$, and $f(E,\mu)=\partial_\mu F(E,\mu)$ is the derivative of the Fermi-Dirac distribution. The density of states $d_i(E)=\frac{1}{2\pi}\Gamma^i/((E-E_i)^2+{(\Gamma^i/2)^2})$ has a Lorentzian broadening due to the finite lifetime caused by the
coupling $\Gamma^i=\Gamma_L |\Psi_L^i|^2 + \Gamma_R |\Psi_R^i|^2$ from the contacts.  
The bias in the occupation in the left and right contacts generates a net external current $J^{ij}=(eU/2\hbar)(\Gamma_L^{ij}-\Gamma_R^{ij})[D^j(\mu)+D^i(\mu)]$ that should be countered by a slight change in the density matrix $\delta\varrho^{ij}=\varrho^{ij}-\varrho^{ij}_{eq}$ of the molecule to achieve a stationary state. In leading order, the change of the stationary density matrix satisfies the equation
$0=\frac{1}{\hbar}\sum_{pq}L^{ijpq}\delta\varrho^{pq}+J^{ij}$, where the operator $L$ is the evolution operator of the density matrix, and the Liouville term is linearized around the equilibrium Fermi-Dirac distribution. In the expression of the linearized operator,
\begin{equation}
    L^{ijpq}=-i(E_i-E_j)\delta_{ip}\delta_{jq}-\frac{1}{2}(\Gamma^{ip}\delta_{jq}+\delta_{ip}\Gamma^{qj})+\frac{1}{2}\left(\tilde{\gamma}^{ip}+\tilde{\gamma}^{jp}\right)\delta_{ij}\delta_{pq} -\frac{1}{2}\sum_r(\tilde{\gamma}^{ri}+\tilde{\gamma}^{rj})\delta_{ip}\delta_{jq},  
\end{equation}
new transition rates appear, which are related to the old ones by $\tilde{\gamma}^{ij}=\gamma^{ij}(1-F(E_i,\mu))/(1-F(E_j,\mu)$.
Finally, the conductance is given in terms of the inverse of the linearized evolution operator, the escape rates of the contacts, and the molecular
orbitals at the contact points
\begin{equation}
    G=-\frac{e^2\Gamma_L\Gamma_R}{\hbar}  
    \sum_{ijpq} \{\Psi_L^i\Psi_L^j[L^{-1}]^{jipq}\Psi^p_R\Psi_R^q
    +\Psi_R^i\Psi_R^j[L^{-1}]^{jipq}\Psi^p_L\Psi_L^q\} D^p(\mu), 
\end{equation}
which is a new formula and our main theoretical result here. An earlier version of this formula with more restrictive approximations has been introduced in Ref.~\cite{papp2019landauer} and has successfully been applied to understand the temperature dependence of the current flowing through protein monolayer junctions~\cite{papp2023experimental}.

\section*{Conductance calculations}

Inverting the $N^2\times N^2$ dimensional matrix of the evolution operator for macromolecules like extracellular cytochrome nanowires with $N\sim 10^4$ atomic orbitals is an elusive task. The eigendecomposition of the inverse of the matrix is dominated by the reciprocal of its smallest eigenvalue, and
in Ref.~\cite{segal2000electron}, it has been shown that in donor-bridge-acceptor molecular systems, it is a good approximation. The 
underlying physical assumption is that the relaxation to equilibrium is faster than the escape of the electrons from the system. 
In \textcolor{blue}{Supplementary Information S3}, we calculated the conductance in this approximation. The result consists of
three terms describing three distinct mechanisms of the total conductance $G=G_{LB}+G_T+G_M$, where
\begin{eqnarray} 
    G_{LB}&=&\frac{2e^2}{h}T,\label{GLB}\\
    G_T&=&\frac{2e^2}{\hbar }\frac{{Z}_L{Z}_R}{{Z}_L+{Z}_R},\label{GT} \\
    G_M&=&\frac{e^2}{h}\left[\frac{{Z}_L}{{Z}_L+{Z}_R}T_{R}+\frac{{Z}_R}{{Z}_L+{Z}_R}T_{L}\right]. \label{GM}
\end{eqnarray}    
 The first term $G_{LB}$ is temperature independent and gives the Landauer-Büttiker formula, where the transmission between the left and right contacts $T=\sum_k \Gamma_L\Gamma_R|\Psi_{L}^k|^2|\Psi_{R}^k|^2/((\mu-E_k)^2+(\Gamma^k/2)^2)$ is in the Breit-Wigner approximation~\cite{lambert2015basic}. It describes coherent elastic processes. This term is suppressed by tunneling in protein structures since the product is exponentially small $|\Psi_{L}^k|^2|\Psi_{R}^k|^2\sim e^{-{l}_{LR}/{l}_T}$, where $l_{LR}$ is the distance of contacts and $l_T\sim 1$ \r{A} is the tunneling length.
The second term $G_T$ describes thermal excitation-based conductance, where the terms ${Z}_{L/R}=\sum_k\Gamma_{L/R}|\Psi_{L/R}^k|^2/4kT\cosh^2((\mu-E_k)/2kT)$ are proportional with the probability for an electron to get from contact $L$ or $R$ excited to one of the levels of the molecule.
The ratio ${Z}_{L/R}/({Z}_L+{Z}_R)$ is the equilibrium probability that an electron from the molecule ends up in contact $L$ or $R$.
The combination ${Z}_{L}{Z}_{R}/({Z}_L+{Z}_R)$ is the probability that an electron from
contact $L$ gets via thermal excitation into the molecule and then from the molecule to contact $R$. 
The third term, $G_M$, is new and describes the mixed process when the electron tunnels from the contact into the molecule described by the transmission $T_{L/R}=\sum_k \Gamma_{L/R}^2|\Psi_{L/R}^k|^4/((\mu-E_k)^2+(\Gamma^k/2)^2)$ and ends up in the other contact via a thermal process. 

\begin{figure*}
\centering
\includegraphics[width=.75\linewidth]{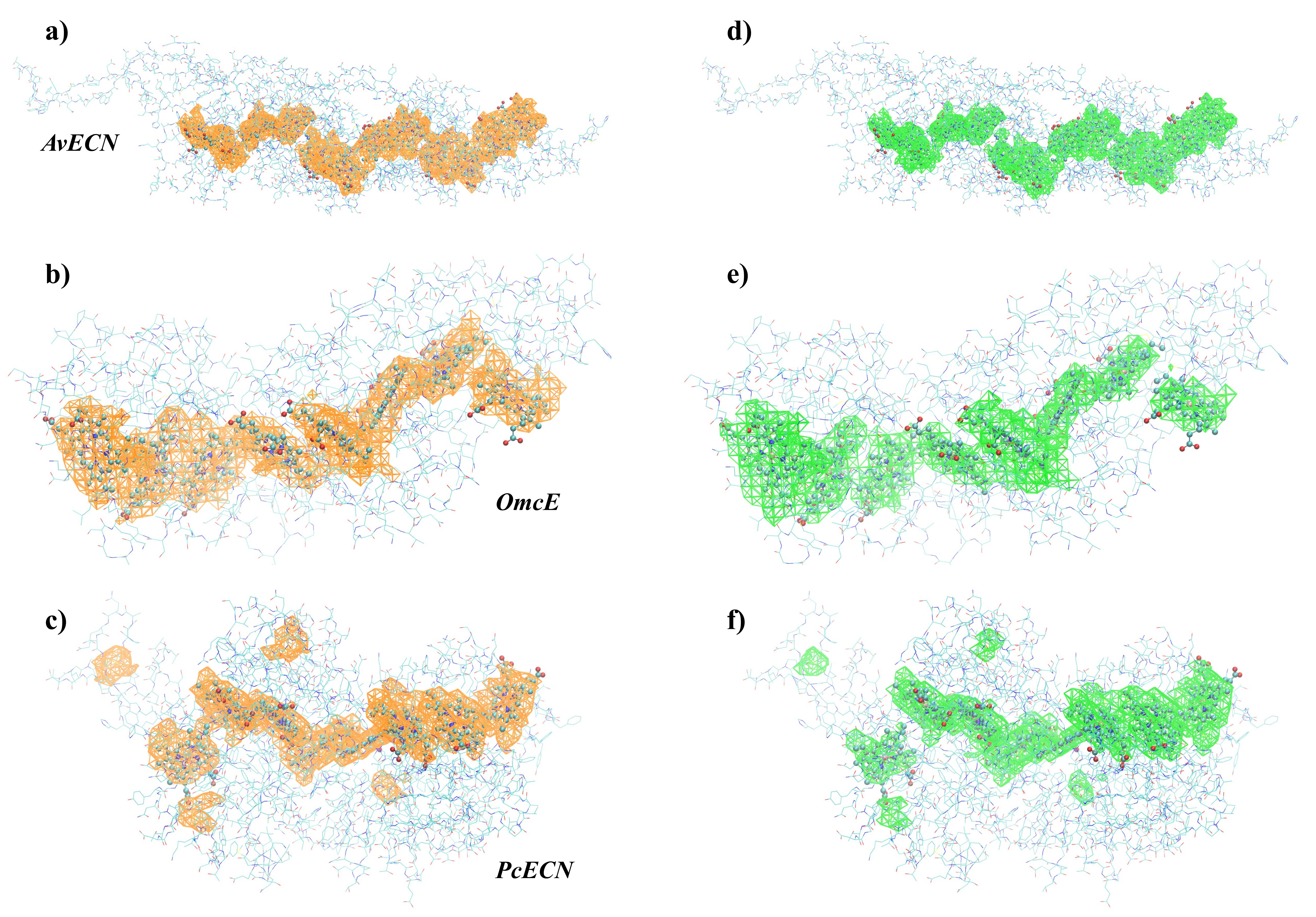}
\caption{Visualization of the functions ${\mathcal{Z}}({\bf r})$ (a-c) and ${\mathcal{T}}({\bf r})$ (d-f). Highly conductive regions of the molecules are located where both of these functions have high values. On (a-c), the orange wireframe meshes show the regions where ${\mathcal{Z}}({\bf r})$ takes on high values in AvECN (PDB ID: \href{https://www.rcsb.org/structure/8E5G}{8E5G}), OmcE (PDB ID: \href{https://www.rcsb.org/structure/7TFS}{7TFS}) and PcECN (PDB ID: \href{https://www.rcsb.org/structure/8E5F}{8E5F}). (d-f) shows the same ECNs as on the left side. The green wireframe meshes indicate the regions where ${\mathcal{T}}({\bf r})$ takes on high values. Lines represent the amino acids, and the heme molecules are depicted using the ball-stick model. Hydrogen atoms are omitted for simplicity.}
\label{fig:z_t}
\end{figure*}

The conductance depends on the position of the contacts and the strengths $\Gamma_L$ and $\Gamma_R$. High conductance is achieved potentially when the contacts are on sites with a high $T$ or $Z$ value. We can visualize the structure by introducing the spatially continuous versions of $T$ and $Z$
\begin{eqnarray}\label{eq:z_t}
    \mathcal{Z}({\bf r})&=&\sum_k \frac{|\Psi^k({\bf r})|^2}{\cosh^2((\mu-E_k)/2k_T)}, \label{Z}\\
    \mathcal{T}({\bf r})&=&\sum_k \frac{|\Psi^k({\bf r})|^4}{(\mu-E_k)^2}, \label{T}
\end{eqnarray}
where $\Psi^k({\bf r})$ are the molecular orbitals in space, and the $\Gamma_{L/R}$
dependence has been removed in order to make them independent of the coupling strengths and the position of a second electrode.
Highly conducting parts of the molecule are those where $\mathcal{Z}({\bf r})$ and $\mathcal{T}({\bf r})$ take on high values.

\section*{Extracellular cytochrome nanowires}

Certain bacterial species can synthesize conductive protein filaments, also known as bacterial nanowires, to facilitate electron export into extracellular environments for respiration purposes and interspecies electron exchange. One of the extensively studied bacteria is \textit{Geobacter sulfurreducens}, a soil bacterium that produces various types of extracellular cytochrome nanowires (ECNs) ~\cite{filman2019cryo,wang2019structure,yalcin2020electric,wang2022cryo} that are composed of cytochrome monomers with either 4 (OmcE), 6 (OmcS), or 8 (OmcZ) hemes placed inside the protein, allowing the bacteria to transport electrons over micrometers. OmcE plays a crucial role in extracellular respiration and is also involved in extracellular conductivity\cite{wang2022cryo}. OmcS is essential at the early stages of biofilm growth, direct electron transfer between co-cultures, and Fe(III) oxide reduction\cite{wang2019structure}. OmcZ nanowires can form a thick conductive biofilm network, possibly because of the branched heme arrangement that leads to one solvent-exposed heme per subunit\cite{10.7554/eLife.81551}.\\
A very recent discovery has brought to light the existence of ECNs in two species of hyperthermophilic archaea, namely Pyrobaculum calidifontis (PcECN) and Archaeoglobus veneficus (AvECN) \cite{baquero2023extracellular}. These ECNs also serve as mediators for long-range extracellular electron transfer. Although the subunits of ECNs don't show similarities in their folds, the hemes' arrangement is common, indicating an evolutionarily optimized structure~\cite{baquero2023extracellular}.
It was previously thought that the metal ions were responsible for the long-range electron transport in heme-containing proteins, but recent research indicates that this process is actually dictated by the porphyrin rings~\cite{agam2020porphyrin}. An experimental study combined with density functional theory calculations on a gold-small tetraheme protein-gold junction also reached the same conclusion~\cite{futera2020coherent}.
While cryo-electron microscopy has enabled the determination of atomic-level structures of these filaments, offering new avenues for theoretical and computational studies, the layout of the conducting and insulating components of ECNs has yet to be determined and visualized. Considering the large size of ECNs, numerical studies that involve entire molecules or oligomers are quite challenging. However, due to their diverse applications~\cite{malvankar2014microbial, liu2020power, bond2003electricity, smith2020bioelectronic, liu2020multifunctional} and similar structure, ECNs are excellent candidates for investigating the underlying mechanism of electron transport across long biological nanowires. Additionally, measurements display a thousand-fold higher conductivity in OmcZ nanowires compared to OmcS ones~\cite{yalcin2020electric}, offering a valuable opportunity to assess how well our calculations capture this significant relative difference.

\begin{figure*}[h]
\centering
\includegraphics[width=.8\linewidth]{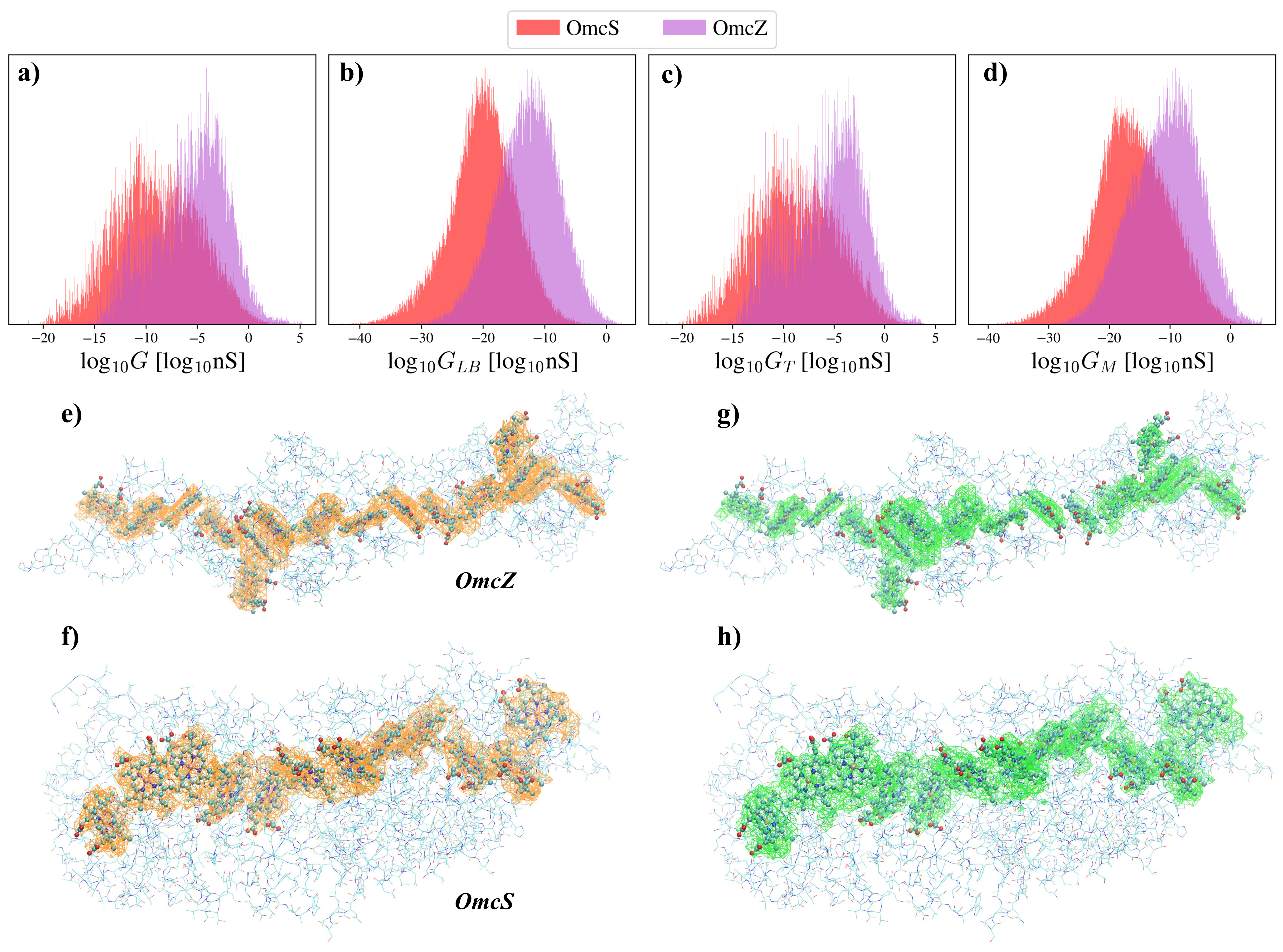}
\caption{Distributions of the logarithm of the conductance $G$ as defined by Eqs.(\ref{GLB}-\ref{GM}) for OmcZ and OmcS, calculated with $\Gamma_R = \Gamma_L = 0.1\,eV$ and $T = 300\,K$ (a-d), and visualization of the functions ${\mathcal{Z}}({\bf r})$ (e,f) and ${\mathcal{T}}({\bf r})$ (g,h) for OmcZ (PDB ID: \href{https://www.rcsb.org/structure/7LQ5}{7LQ5}) and OmcS (PDB ID: \href{https://www.rcsb.org/structure/6ef8}{6EF8}). The total conductance $G$ and all its terms follow a lognormal-like distribution. For both structures, the Landauer-Büttiker term $G_{LB}$ of the conductance (b) is the smallest, the dominating term is the thermal excitation-based conductance $G_T$ (c), and the last term $G_M$ (d) that describes a mixed process, is in between the two. 
There are several orders of magnitude differences between the means of the total conductance $G$ distributions, aligning with experimental observations that show significantly higher values for OmcZ compared to OmcS~\cite{yalcin2020electric}. 
On (e, f), the orange wireframe meshes show the regions where ${\mathcal{Z}}({\bf r})$ takes on high values in OmcZ and OmcS. On (g, h), the green wireframe meshes show the regions where ${\mathcal{T}}({\bf r})$ takes on high values in OmcZ and OmcS, respectively. Highly conductive regions are located where both functions have high values. Lines represent the amino acids, and the heme molecules are depicted using the ball-stick model. Hydrogen atoms are omitted for simplicity.}
\label{fig:omcs_omcz}
\end{figure*}

Employing Eqs.(\ref{Z}) and (\ref{T}), we can visualize the highly conductive regions within the five ECNs introduced here. 
To represent these regions, we constructed a 3D rectangular grid with $1.5$ \r{A} resolution for each ECN structure and computed the values of the functions ${\mathcal{Z}}({\bf r})$ and ${\mathcal{T}}({\bf r})$ on the grid points at room temperature $kT = 25\,meV$. (See details of the calculations in Methods.) Subsequently, we selected all non-zero values for each function and determined the highest five percent for the function ${\mathcal{Z}}({\bf r})$ and the highest twenty percent for the function ${\mathcal{T}}({\bf r})$. These percentiles serve as the isovalues for the isosurfaces presented in Fig. \ref{fig:z_t} and Fig. \ref{fig:omcs_omcz} (e-h).\\
Consistent resemblances in both ${\mathcal{Z}}({\bf r})$ and ${\mathcal{T}}({\bf r})$ for all ECNs supports the existence of a common mechanism governing long-range electron transport~\cite{baquero2023extracellular}. ${\mathcal{Z}}({\bf r})$ and ${\mathcal{T}}({\bf r})$ have similar structures, with the most conductive parts located inside the proteins, spanning across the porphyrin rings, in agreement with previous studies of heme-containing proteins~\cite{agam2020porphyrin, futera2020coherent}.
In the case of PcECN (Fig. \ref{fig:z_t} c and f), one can also observe highly conductive regions on the intra-subunit disulfide bonds. The functionality of these additional high-conducting areas is an open question.

Our model also allows us to calculate the conductance between any two atomic orbitals of the ECNs, facilitating a comparison of the conductance values for OmcZ and OmcS. Utilizing Eqs.(\ref{GLB}-\ref{GM}), we computed each term ($G_{LB}$, $G_{T}$ and $G_M$) of the total conductance $G$ between $100,000$ pairs of randomly selected atomic orbitals for both OmcZ and OmcS at $T=300\,K$ with parameters $\Gamma_L = \Gamma_R = 0.1\,eV$. In addition, we only considered orbitals that are $30-35$ \r{A} apart, and do not belong to hydrogen atoms. We show the distributions of the calculated values of $G$, $G_{LB}$, $G_{T}$ and $G_M$ for both OmcS and OmcZ in Fig. \ref{fig:omcs_omcz} (a-d). Notably, all of the distributions are lognormal-like.\\
While a detailed quantitative comparison between experimentally measured and computed values is not within the scope of this paper, our calculations reveal several orders of magnitude differences in the conductance of OmcZ and OmcS, consistent with the experimental observation of significantly higher values for OmcZ compared to OmcS.
The potential explanation for this difference in the level of structure lies in the distinct heme-heme interactions due to the closer stacking of the hemes in OmcZ~\cite{yalcin2020electric}. Given the significant reliance of our calculations on individual structures, we arrive at a similar conclusion, that these structural differences are the primary contributors to the observed disparity.

\section*{Distance- and temperature dependence}
\begin{figure*}[ht]
    \centering
    \includegraphics[width=.99\linewidth]{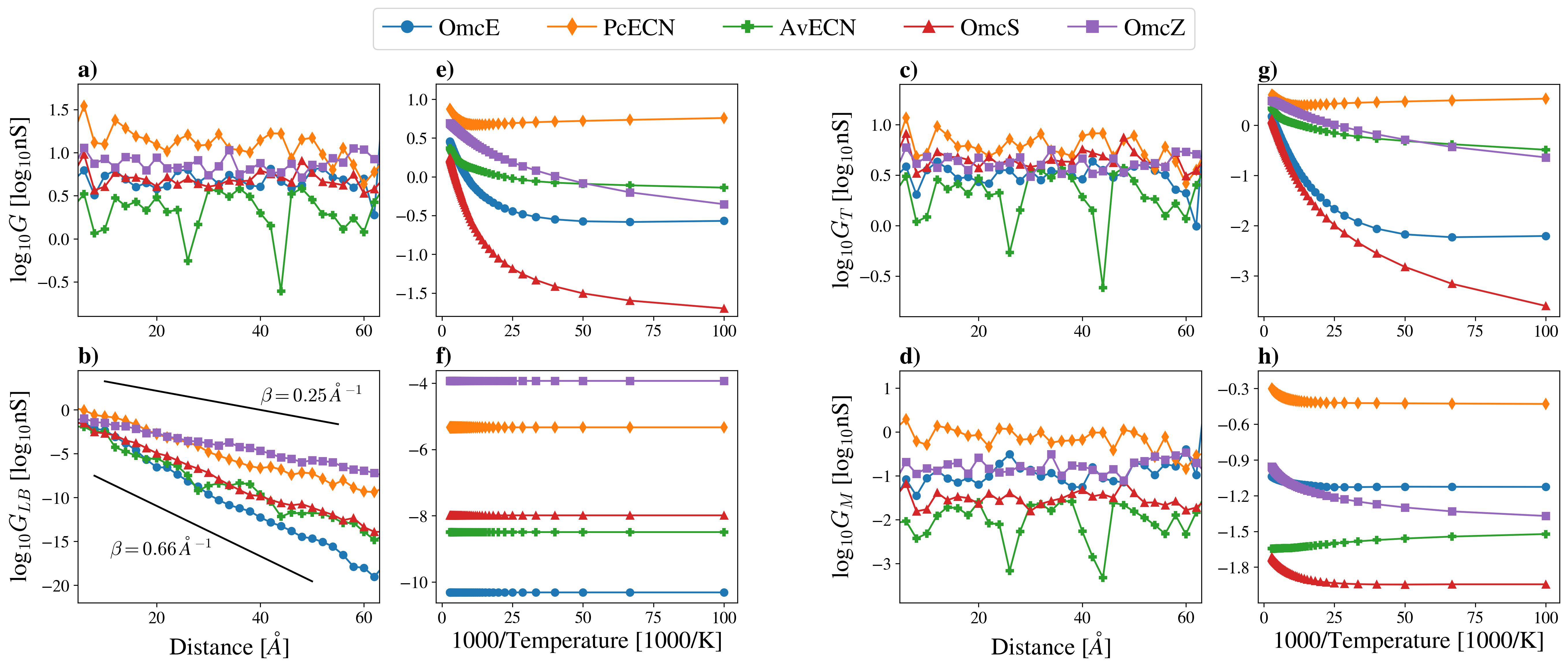}
    \caption{Distance- and temperature dependence of the conductance calculated between atomic orbitals of the ECNs, with coupling strengths $\Gamma_L=\Gamma_R=0.1\, eV$. (a-d) shows the distance dependence of the total conductance $G$, the Landauer-Büttiker term $G_{LB}$, the thermal term $G_T$ and the mixed term $G_M$. On (e-h), the temperature dependence of the conductance is presented. On (b), the two black lines serve as visual guides and follow the form $exp(-\beta d)$, where $d$ is the distance between the two atomic orbitals and $1/\beta$ is the decay length of the conductance.}
    \label{fig:d-t}
\end{figure*}

Using Eqs.\ref{GLB}-\ref{GM}, one can study the temperature and distance dependencies of conductance by evaluating it between atomic orbitals positioned at specified distances from each other and at varying temperatures. (See details of the calculations in Methods.) To investigate the distance dependence, the conductance was computed between $400,000$ randomly selected pairs of atomic orbitals belonging to the hemes, which are the most highly conducting parts of the ECNs. The calculation excluded atomic orbitals associated with hydrogen atoms. Then, the mean of the logarithmic conductance values was calculated for each specific distance with a tolerance of $\pm 0.1$ \r{A}. The results are presented in Fig. \ref{fig:d-t}(a-d).
As Fig. \ref{fig:d-t}(a) shows, the total conductance $G$ fluctuates in the $0.1-1\,nS$ range without a clear decreasing trend. A similar pattern is also observed for $G_T$ and $G_M$ fluctuating in various ranges. 
This reflects the fact that the ratio ${Z}_{L/R}/({Z}_L+{Z}_R)$ associated with the thermal exit probability does not depend on the distance on average.
The $G_{LB}$ conductance, which is associated with pure tunneling, decays exponentially with the distance $d$ between the two atomic orbitals $G_{LB}\propto exp(-\beta d)$, where the distance decay constant $\beta$ $\approx 0.25-0.66$ \r{A}$^{-1}$. In Ref.~\cite{futera2023tunneling}, other multiheme cytochrome proteins were investigated, demonstrating a similar result with $\beta = 0.2$ \r{A}$^{-1}$.

Concerning the temperature dependence, a set of $10,000$ pairs of randomly selected atomic orbitals of atoms from the hemes, positioned at distances within the range of $30-35$ \r{A}, was considered. Atomic orbitals belonging to hydrogen atoms were excluded from this analysis. Conductance values between these pairs were computed across a temperature range of $10-360\,K$. The mean of the obtained logarithmic values was calculated for each temperature as shown in Fig. \ref{fig:d-t}(e-h). At high temperatures, the transport is clearly thermally activated, and the term $G_T$ dominates the conductance for all ECNs presented here at high temperatures. Interestingly, the dominance shifts to the mixed term $G_M$ in the case of OmcS and OmcE below temperatures $\approx 42\,K$ and $\approx 67\,K$, respectively. As expected, the Landauer-Büttiker term $G_{LB}$ is temperature independent. The mixed tunneling-thermal term $G_M$ falls off with temperature considerably slower than $G_T$ and reaches a finite plateau value at very low temperatures. This is in line with experimental results for several other proteins~\cite{sepunaru2011solid, bera2023near, papp2023experimental, kayser2019solid, garg2018interface, fereiro2019solid}, whereupon cooling, the current through a protein layer exponentially decreases at high temperatures and then reaches a constant value at low temperatures.

In summary, through the integration of the Liouville-Master Equation's density matrix and a phenomenological model of electronic conductance in molecular systems, we successfully computed and visually represented the conductance attributes of ECNs. Our findings reveal that ECNs resemble insulated cables, characterized by a highly conductive inner core spanning the chain of porphyrin rings within the proteins. This observation aligns with recent experimental findings emphasizing the significance of porphyrin rings within these filaments~\cite{agam2020porphyrin} and lends support to the proposition that ECN structures have undergone evolutionary optimization for an optimal heme arrangement~\cite{baquero2023extracellular}. We explored the conductance's and its components' distance and temperature dependence, finding that the conductance of the studied ECNs falls within the range of $0.1-1\,nS$ with fluctuations but lacks a clear decrease with distance. However, the Landauer-Büttiker term of the total conductance exhibits exponential decay with a decay length of $1/\beta \approx 1.5-4$ \r{A}, resembling the decay length of $1/\beta = 5$ \r{A} observed in other multiheme cytochromes~\cite{futera2023tunneling}. Moreover, the simulated temperature-dependence of the conductance in the studied ECNs is consistent with the typical behavior displayed by other proteins~\cite{sepunaru2011solid, bera2023near, papp2023experimental, kayser2019solid, garg2018interface, fereiro2019solid}. 

Our newly developed formula possesses the capability to model the conductance of diverse proteins, thereby contributing significantly to the field of nano-bioelectronics. Notably, this formulation incorporates the electron-phonon interaction in the intermediate temperature range, addressing a critical aspect given the non-isolated nature of proteins. This approach enhances the applicability and relevance of our results in advancing the understanding and application of protein electron transport in varied bioelectronic contexts.

\section*{Methods}

\subsection*{Structure preparation}
First, we downloaded the cryo-EM structures from the Protein Data Bank (PDB) with PDB IDs \href{https://www.rcsb.org/structure/7TFS}{7TFS}, \href{https://www.rcsb.org/structure/6ef8}{6EF8}, \href{https://www.rcsb.org/structure/7LQ5}{7LQ5}, \href{https://www.rcsb.org/structure/8E5G}{8E5G} and \href{https://www.rcsb.org/structure/8E5F}{8E5F} for OmcE\cite{wang2022cryo}, OmcS\cite{wang2019structure}, OmcZ\cite{gu2023structure}, AvECN\cite{baquero2023extracellular} and PcECN\cite{baquero2023extracellular}, respectively. We created dimers and a trimer from the biological assemblies in the case of AvECN. In Maestro (\textit{Schrödinger Release 2023-4: Maestro, Schrödinger, LLC, New York, NY, 2023.}, \href{https://www.schrodinger.com/platform/products/maestro/}{https://www.schrodinger.com/platform/products/maestro/}), we connected the appropriate cysteines to the hemes and conducted a force-field minimization procedure only on the side chains of those cysteines to optimize the spatial arrangement of the atoms. Then, C-terminal oxygen atoms and missing hydrogen atoms were added to the protein structures.

\subsection*{Quantum chemistry calculations}
 After preparing the structures, the Hamiltonian and overlap matrices were calculated. Due to the large size of the proteins, we opted for the Extended Hückel method instead of the more common DFT calculations, which can consider the redox state of each cofactor. Nevertheless, we took the redox states into account by considering the total charge of the protein. The YAeHMOP software (version 3.0.1, \href{https://yaehmop.sourceforge.net/}{https://yaehmop.sourceforge.net}) was used for performing the extended Hückel calculations on the molecules, and it requires the positions of the atoms as an input. 
 In the semi-empirical extended Hückel method, the total valence electron wavefunction is described as the product of one-electron wavefunctions:
 \begin{equation}
     \Phi_{valence}=\Psi^1(1)\Psi^2(2)\Psi^3(3)...\Psi^k(n),
 \end{equation}
 where $k$ and $n$ denote the molecular orbital and the number of the electron, respectively. Each molecular orbital is constructed by Linear Combination of Atomic Orbitals (LCAO):
 \begin{equation} \label{lcao}
     \Psi^k=\sum_{r=1}^NW_{r}^{k}\varphi_r.\qquad k=1,2,3,...N.
 \end{equation}
Here $\varphi_r$-s are the valence electrons' Slater-type atomic orbitals, that form the basis set. The $W_{r}^{k}$ coefficient is the weight of the $r^{th}$ atomic orbital in the $k^{th}$ molecular orbital.
To calculate the coefficients and the spectrum of a molecule, one needs to solve the following generalized eigenvalue problem:
\begin{equation}
    HW=ESW,
\end{equation}
where $S$ is the overlap matrix, $H$ is the Hamiltonian matrix~\cite{lowe2011quantum}, $E$ is the diagonal matrix of molecular orbital energies (energy eigenvalues), and $W$ is the matrix of the eigenvectors containing the linear combination coefficients used in the LCAO method.
The Hamiltonian and overlap matrices are the outputs of YAeHMOP, and the eigenvalue problem was solved using the SciPy package~\cite{2020SciPy-NMeth} in Python.
The Highest Occupied Molecular Orbital (HOMO) and the Lowest Unoccupied Molecular Orbital (LUMO) for all structures have been determined automatically based on the total charge of the molecule calculated by the Maestro software (\textit{Schrödinger Release 2023-4: Maestro, Schrödinger, LLC, New York, NY, 2023.}, \href{https://www.schrodinger.com/platform/products/maestro/}{https://www.schrodinger.com/platform/products/maestro/}) at neutral pH. In the case of the OmcS dimer, the HOMO level has been positioned two levels above the automatic value due to the presence of two redundant hydrogen atoms within the structure.
 
\subsection*{Visualization}
To visualize the coupling strength-independent functions ${\mathcal{Z}}({\bf r})$ and ${\mathcal{T}}({\bf r})$, it is sufficient to consider molecular orbitals only within $5\,kT$ from HOMO and LUMO. Molecular orbitals were calculated with the presented LCAO method in Python on grids with $1.5$ \r{A} resolution. We used the gridData module of the MDAnalysis package~\cite{oliver_beckstein_2022_6582343} in Python to write input files for the Visual Molecular Dynamics (VMD) molecular visualization program~\cite{HUMP96, STON1998}.

\subsection*{Conductance calculation}
To calculate the conductance $G$, we performed a Löwdin orthogonalization $\tilde{H} = S^{-1/2}HS^{-1/2}$ first. Subsequently, we computed the conductance between any pair of atomic orbitals in the Löwdin basis using Eqs.(\ref{GLB}-\ref{GM}), where $\Psi_{L/R}^k$ is the $k^{th}$ molecular orbital at the Löwdin atomic orbital $L/R$. To enhance precision, we considered a larger set of molecular orbitals than utilized in the visualization process, specifically $\pm 20$ energy levels above and below the HOMO and LUMO. Molecular orbitals beyond this range have negligible influence on $T$, $T_{L/R}$, and $Z_{L/R}$.\\

\noindent A code workflow chart is provided in the \textcolor{blue}{Supplementary Information S4}. 

\section*{Acknowledgement}
This research was supported by the Hungarian National Research, Development and Innovation Office within the Quantum Information National Laboratory of Hungary (Grant No. 2022-2.1.1-NL-2022-00004).
E.P. received sponsorship from the Gedeon Richter Talentum Foundation in the framework of the Gedeon Richter Excellence Ph.D. Scholarship of Gedeon Richter. 
The authors thank Dóra K. Menyhárd for her valuable assistance during structure preparation. We acknowledge her affiliation at HUN-REN-ELTE Protein Modeling Research Group ELTE Eötvös Loránd University, supported by project no. 2018-1.2.1-NKP-2018-00005 (HunProtExc) of the NRDI Office, financed under the 2018-1.2.1-NKP funding scheme.

\section*{Author contributions}
G.V. designed and directed the project. E.P. conducted the simulations and generated visualizations. All authors contributed in preparing the manuscript.

\section*{Data availability}
The data generated and analyzed during the current study are available from the corresponding author upon request.

\section*{Code availability}
The codes that support the results within this paper are available from the corresponding author
upon request.

\section*{Competing interests}
The authors declare no competing interests.

\newpage

\renewcommand{\appendixname}{Supplementary Information}

\renewcommand{\thesection}{\arabic{section}}

\appendix

\renewcommand{\thefigure}{S\arabic{figure}}
\setcounter{figure}{0}
\renewcommand{\thetable}{S\arabic{table}}
\setcounter{table}{0}
\setcounter{equation}{0}
\renewcommand{\theequation}{S\arabic{equation}}

\begin{center}
    {\Huge \bf Supplementary Information}
\end{center}
\vspace{0.3cm}

\section*{S1. Master equation for electron transport in molecules}

Following Refs.\cite{davis1997electron,segal2000electron}, the reduced density matrix $\varrho$ of an electron in a molecule can be described with the quantum master equation
\begin{equation}
    \partial_t\varrho=\frac{1}{i\hbar}[H,\varrho]+R(\varrho),
\end{equation}
where $H$ is the Hamiltonian of the molecule and
$R(\varrho)$ is an operator for the phenomenological description of the interaction with the environmental heat bath. Electrons can also escape from the molecule. This can be taken into account by adding an imaginary part to the site energy $H_{mm}-i\Gamma_m/2$ so that the escape rate from the site $m$ is $\Gamma_m/\hbar$. Electrons can also be added to a site with a rate $J_m$. The quantum master equation
including escape and currents, is 
\begin{equation}
    \partial_t\varrho=\frac{1}{i\hbar}[H,\varrho]-\frac{1}{\hbar}\{\Gamma,\varrho\}+R(\varrho)+J,
\end{equation}
where the matrix $\Gamma=\text{diag}\{...,\Gamma_m/2,... \}$ and $J$ is
a vector with elements $J_m$.
In site representation, this equation takes the form
\begin{eqnarray}
    \dot{\varrho}_{nm}&=&-\frac{i}{\hbar}\sum_k (H_{nk}\varrho_{km}-\varrho_{nk}H_{km})-\frac{1}{2\hbar}(\Gamma_n+\Gamma_m)\varrho_{nm} +R_{nm}(\varrho)+\delta_{nm}J_n,
\end{eqnarray}
where $R_{nm}(\varrho)$ are the matrix elements of $R(\varrho)$ and summations go for all $N$ sites. The site-based quantum master equation can be transformed into energy representation $\hat{\varrho}$ 
with transformed matrix elements $\varrho^{ij}=\sum_{nm}\Psi_n^i\varrho_{nm}\Psi^j_m$, where $\Psi_n^i$ is
the real eigenfunction $E_i\Psi_n^i=\sum_m H_{nm}\Psi_m^i$ of the Hamiltonian.
The transformed equation is
\begin{eqnarray}
    \dot{\varrho}^{ij}&=&-\frac{i}{\hbar}(E_i-E_j)\varrho^{ij}- \frac{1}{2\hbar}\sum_r (\Gamma^{ir}\varrho^{rj}+\varrho^{ir}\Gamma^{rj}) +R^{ij}(\varrho)+J^{ij},\label{Senergyrep}
\end{eqnarray}                                                             
where $\Gamma^{ij}=\sum_n\Psi_n^i\Gamma_n\Psi_n^j$, $J^{ij}=\sum_n\Psi_n^iJ_n\Psi_n^j$ and $R^{ij}=\sum_{nmkl}\Psi_n^i\Psi_m^jR_{nm}(\varrho)$. 
We assumed that the Hamiltonian is an $N\times N$ real symmetric matrix with real eigenvalues and eigenvectors. 

The operator $R(\varrho)$ should describe the interaction with the environment and ensure the correct equilibrium properties of the electron 
distribution. In the framework of the single electron picture, the following Liouville master equation\cite{gebauer2004kinetic,gebauer2006kohn} has been introduced to describe electron transport in nanoscale systems
\begin{eqnarray}
    R^{ij}(\varrho)&=&(\delta_{ij}-\varrho^{ij})\sum_p \frac{1}{2\hbar}(\gamma^{ip}+\gamma^{jp})\varrho^{pp}
    -\varrho^{ij}\sum_p \frac{1}{2\hbar}(\gamma^{pi}+\gamma^{pj})(1-\varrho^{pp}),\label{SCar}
\end{eqnarray}
which can account for the exclusion principle, and in equilibrium, in the absence of external currents and escape, it leads to the Fermi distribution
$\varrho^{ij}_{eq}=\delta_{ij}/(1+e^{(E_i-\mu)/kT})$, where $\mu$ is the chemical potential of the system. For molecules, the electrons are coupled to the phonon bath, which induces transitions between the energy levels with rates
\begin{equation}
    \gamma^{ij}=\gamma(\omega_{ij}) \sum_{n}|\Psi_n^i|^2|\Psi_n^j|^2,
\end{equation}
where $\omega_{ij}=(E_i-E_j)/\hbar$ is the transition frequency and $\gamma(\omega)$ is the spectral density of the bath, and 
obeys the Boltzmann-type detailed balance equation 
\begin{equation}
\gamma^{ij}/\gamma^{ji}=e^{-(E_i-E_j)/kT}. \label{detailed}
\end{equation}
There are several 
models used for the approximation of the spectral density for proteins.
One of them is the Ohmic oscillator bath with cutoff\cite{mohseni2008environment}
\begin{equation}
    \gamma(\omega)=\eta\frac{\hbar\omega}{e^{\hbar\omega/kT}-1}e^{-|\omega|/\omega_c},
\end{equation}
where $\hbar\omega_c\approx 0.0185 eV$ is the typical cutoff frequency for proteins. The parameter $\eta=2\pi E_R/\hbar\omega_c\approx 1.46$ where $E_R$ is the reorganization energy.

\section*{S2. Conductance formula}

Our starting point is the derivation of Datta et al.'s low-temperature conductance formula for molecules in Refs.\cite{zahid2003electrical,datta1997electronic}, which we summarize here briefly. The molecule is coupled to a left and a right electrode.
The discrete levels of the molecule $E_i$ are non-resonantly coupled to the left
and right electrode sites with indices $n=L$ and $n=R$  with coupling strengths $\Gamma_L$ and $\Gamma_R$, respectively. The presence of contacts
broadens the level density, and the energy levels became resonances $E_i-i\Gamma^i/2$,
where $E_i$ and $\Gamma^i$ are given by the real and imaginary parts of the eigenvalues of the
Hamiltonian $H_{nm}-i\delta_{nm}\Gamma_n/2$. The local density of states has a Lorentzian form
\begin{equation}
d_i(E)=\frac{1}{2\pi}\frac{\Gamma^i}{(E-E_i)^2+{(\Gamma^i/2)^2}}.
\end{equation}
For concreteness, here we assume tactically that the contacts can be taken into account perturbatively and 
in the first order the real parts $E_i$ are unperturbed, while $\Gamma^i=\Gamma_L |\Psi_L^i|^2 + \Gamma_R |\Psi_R^i|^2$.

 If the level $E_i$ is in equilibrium with a contact, then its occupation is
\begin{equation}
\varrho^i_{eq}(\mu)=2\int^{+\infty}_{-\infty} dE d_i(E)F(E,\mu),
\end{equation}
where $\mu$ is the chemical potential in the contact, $F(E,\mu)=(1+e^{(E-\mu)/kT})^{-1}$ is the Fermi distribution, and the factor 2 stands for spin degeneracy. The reduced density matrix of the system in equilibrium $\hat{\varrho}_{eq}$ is diagonal
$\varrho^{ij}_{eq}=\delta_{ij}\varrho^i_{eq}(\mu)$. If the occupation of the levels differs from the equilibrium with the left contact or with the right contact, then it generates material current out of the
left and right contacts 
\begin{eqnarray}
    J_L&=&\frac{1}{\hbar}\text{Tr}\{\hat{\Gamma}_L(\hat{\varrho}_{eq}(\mu_L)-\hat{\varrho})\},  \\
    J_R&=&\frac{1}{\hbar}\text{Tr}\{\hat{\Gamma}_R(\hat{\varrho}_{eq}(\mu_R)-\hat{\varrho})\}, 
\end{eqnarray}
where $\mu_{L/R}$ are the chemical potentials of the contacts and the operators
$\hat{\Gamma}_{L/R}$ have matrix elements $\Gamma^{ij}_{L}=\Gamma_{L}\Psi_{L}^i\Psi_L^j$ and $\Gamma^{ij}_{R}=\Gamma_{R}\Psi_{R}^i\Psi_R^j$.
The two currents are equal and opposite in sign $J_L=-J_R=J$. The electric current is proportional to the material current $I=eJ$. When the electric field is switched on ($U\neq 0$), the chemical potentials at the left and the right contacts differ $\mu_{L/R}=\mu\pm eU/2$, and a net electric current flows.
In the linear regime, we can expand the deviation of level occupations from their equilibrium values in the leading order
\begin{eqnarray}
\varrho_{eq}^i(\mu_{L/R})&\approx&2\int_{-\infty}^{+\infty}dEd_i(E)[F(E,\mu)\pm f(E,\mu)eU/2 ]  =\varrho_{eq}^i(\mu) \pm D^i(\mu)eU,
\end{eqnarray}
where $D^i(\mu)=\int_{-\infty}^{+\infty} f(E,\mu)d_i(E)dE$ and $f(E,\mu)=\partial_\mu F(E,\mu)=1/4kT\cosh^2((\mu-E)/2kT)$ is
the derivative of the Fermi distribution. In molecules, the equilibrium chemical potential is typically between the highest occupied (HOMO) and lowest unoccupied (LUMO) molecular orbitals $E_{HOMO}<\mu<E_{LUMO}$. The derivative of the
Fermi distribution is strongly peaked at $E=\mu$ and decays quickly 
far from the peak.
The Lorentzian density of states is strongly peaked at $E=E_i$. Combining these delta function-like behaviors yield
\begin{equation}
    D^i(\mu)\approx d_i(\mu)+f(E_i,\mu),\label{S14}
\end{equation}
where the first term describes electron tunneling, and the second is the thermal excitation.  

The population of the levels in the molecule shift from their equilibrium values. We can introduce the deviation $\delta\hat{\varrho}=\hat{\varrho}-\hat{\varrho}_{eq}$. In leading order, the current can be written as
\begin{eqnarray}
    J_L&=&\frac{1}{\hbar}\operatorname{Tr}(\hat{\Gamma}_L(\hat{D}eU-\delta\hat{\varrho})), \label{Lcurrent} \\ 
    J_R&=&-\frac{1}{\hbar}\operatorname{Tr}(\hat{\Gamma}_R(\hat{D}eU+\delta\hat{\varrho})), 
\end{eqnarray}
where the equilibrium value of the density matrix dropped out, and only deviations from the equilibrium remain, and
$\hat{D}$ is the operator with matrix elements $D^{ij}=\delta_{ij}D^i(\mu)$.
The deviation $\delta\varrho^i$ satisfies the stationary version of the master equation (\ref{energyrep})
\begin{eqnarray}
0&=&-\frac{i}{\hbar}(E_i-E_j)\delta\varrho^{ij}+\delta R^{ij}
-\frac{1}{2\hbar} \sum_r (\Gamma^{ir}\delta\varrho^{rj}+\delta\varrho^{ir}\Gamma^{rj})
+\frac{eU}{2\hbar}(\Gamma_L^{ij}-\Gamma_R^{ij})(D^j+D^i), \label{stacio}
\end{eqnarray}    
where $\delta R^{ij}=R^{ij}(\hat{\varrho}_{eq}+\delta\hat{\varrho})-R^{ij}(\hat{\varrho}_{eq})$ and we have to keep the terms linear in the deviation only.
The linearized expression takes the form
\begin{eqnarray}
 \delta R^{ij}  &=& \frac{\delta_{ij}}{2\hbar}\sum_p\left(\frac{1-F_i}{1-F_p} \gamma^{ip}+\frac{1-F_j}{1-F_p} \gamma^{jp}\right)\delta\varrho^{pp} 
 -\frac{\delta\varrho_{ij}}{2\hbar}\sum_p\left(\frac{1-F_p}{1-F_i}\gamma^{pi}+\frac{1-F_p}{1-F_j}\gamma^{pj}\right),\label{eq18}
\end{eqnarray}
where $F_i=(1+e^{(E_i-\mu)/kT})^{-1}$ is the Fermi distribution. We can introduce new transition rates with the
definition $\tilde{\gamma}^{ij}=\gamma^{ij}(1-F_i)/(1-F_j),$
and then (\ref{eq18}) takes a simpler form
\begin{equation}
 \delta R^{ij}  = \frac{\delta_{ij}}{2\hbar}\sum_p\left(\tilde{\gamma}^{ip}+\tilde{\gamma}^{jp}\right)\delta\varrho^{pp} -\frac{\delta\varrho_{ij}}{2\hbar}\sum_p\left(\tilde{\gamma}^{pi}+\tilde{\gamma}^{pj}\right).
\end{equation}
The new transition rates change the detailed balance (\ref{detailed}) to
\begin{equation}
\tilde{\gamma}^{ij}/\tilde{\gamma}^{ji}=\frac{(1-F_i)^2}{(1-F_j)^2}e^{-(E_i-E_j)/kT}=\frac{\cosh^2((E_j-\mu)/2kT)}{\cosh^2((E_i-\mu)/2kT)},\label{db}
\end{equation}
which now reflects the Fermi statistics. If the magnitude of the difference between the chemical potential and the energy is much larger than the thermal energy, 
then $\cosh^2((E_i-\mu)/2kT)\approx e^{|E_i-\mu|/kT}/4$ and
\begin{equation}
    \tilde{\gamma}^{ij}/\tilde{\gamma}^{ji}\approx e^{-(|E_i-\mu|-|E_j-\mu|)/kT},
\end{equation}
is the Boltzmann-type detailed balance equation
reflecting that states below the chemical potential can be regarded as hole states, and $|E_i-\mu|$ is the energy of the hole. 

Finally, we can introduce the evolution operator
\begin{eqnarray}
    L^{ijpq}&=&-i(E_i-E_j)\delta_{ip}\delta_{jq}-\frac{1}{2}(\Gamma^{ip}\delta_{jq}+\delta_{ip}\Gamma^{qj})  
    +\frac{1}{2}\left(\tilde{\gamma}^{ip}+\tilde{\gamma}^{jp}\right)\delta_{ij}\delta_{pq} -\frac{1}{2}\sum_r(\tilde{\gamma}^{ri}+\tilde{\gamma}^{rj})\delta_{ip}\delta_{jq}, \nonumber \\
\end{eqnarray}
and write (\ref{eq18}) as
\begin{equation}
    0=\frac{1}{\hbar}\sum_{pq}L^{ijpq}\delta\varrho^{pq}+J^{ij}\label{evolution},
\end{equation} 
where $\Gamma^{ij}=\Gamma^{ij}_L+\Gamma^{ij}_R$ and $J^{ij}=(eU/2\hbar)(\Gamma_L^{ij}-\Gamma_R^{ij})(D^j+D^i)$. We can solve (\ref{evolution}) for the density deviation yielding
\begin{equation}
    \delta \varrho^{ij}=\frac{eU}{2}\sum_{pq}[L^{-1}]^{ijpq}(\Gamma_R^{pq}-\Gamma_L^{pq})(D^q+D^p).\label{rhodelta}
\end{equation}
We can substitute this back to (\ref{Lcurrent}) and get the material current $J_L$ and electric current $I=eJ_L$ trough the molecule
\begin{eqnarray}
    I=\frac{e^2U}{\hbar}  \left[\sum_i\Gamma^{ii}_LD^i 
    - \frac{1}{2}\sum_{ijpq} \Gamma_L^{ij}[L^{-1}]^{jipq}(\Gamma_R^{pq}-\Gamma_L^{pq})(D^q+D^p)\right]
\end{eqnarray}

Probability conservation guarantees that in the absence of external currents
the initial probability content of a density matrix 
is leaving the system through the contacts. Assume that we start with $\varrho^{ij}(0)=\delta_{ip}\delta_{jq}$. The material current flowing out through the contacts is given by
\begin{equation}
j(t)=\frac{1}{\hbar}\sum_{ij}\Gamma^{ji}\varrho^{ij}(t),\label{Jcurr}
\end{equation}
The total probability through the contacts is equal to the initial probability $\int_0^{\infty}j(t)dt=\operatorname{Tr}(\varrho(0))=\delta_{pq}$.
The evolution equation in the absence of external current is
\begin{equation}
    \partial_t\varrho^{ij}=\frac{1}{\hbar}\sum_{kl}L^{ijkl}\varrho^{kl},\label{eveq}
\end{equation} 
and its solution is $\varrho^{ij}(t)=\sum_{kl}\left[\exp(Lt/\hbar)\right]^{ijkl}\varrho^{kl}(0)$. Substituting this solution into (\ref{Jcurr}) and integrating in time yields
\begin{equation}
-\sum_{ij}\Gamma^{ji}\left[ L^{-1}\right]^{ijpq}=-\sum_{ij}(\Gamma^{ij}_L+\Gamma^{ij}_R)[L^{-1}]^{ijpq}=\delta_{pq}.\label{identity}
\end{equation}
Using this relation, we can write the conductance $G=I/U$ in a manifestly symmetric form
\begin{eqnarray}
    G&=&-\frac{e^2\Gamma_L\Gamma_R}{\hbar}  
    \sum_{ijpq} \{\Psi_L^i\Psi_L^j[L^{-1}]^{jipq}\Psi^p_R\Psi_R^q 
    +\Psi_R^i\Psi_R^j[L^{-1}]^{jipq}\Psi^p_L\Psi_L^q\} D^p(\mu), \label{cndf}
\end{eqnarray}
which is the main result and allows the calculation of the conductance in terms of the resolvent (\ref{evolution}) 
of the evolution operator.

\section*{S3. Perturbative treatment}

Numerical inversion of the $N^2\times N^2$ matrix of the evolution operator is a daunting task, 
therefore, we develop a perturbative approximation that significantly reduces the computational effort and
allows an analytic treatment if certain conditions are met.
We can introduce the left and the right eigenfunctions of the evolution operator
\begin{eqnarray}
    \lambda_kV^{ij}_k&=&\sum_{pq}L^{ijpq}V^{pq}_k,\\
    \lambda_kU^{ij}_k&=&\sum_{pq}L^{pqij}U^{pq}_k.
\end{eqnarray}
The inverse of the operator can be expressed in terms of the eigenfunctions and eigenvalues
\begin{equation}
    [L^{-1}]^{ijpq}=\sum_{k=0}^\infty\frac{1}{\lambda_k}V^{ij}_kU^{pq}_k,
\end{equation}
so that terms with nearly zero $\lambda\approx 0$ eigenvalues dominate the inverse. 
For a closed system $\Gamma^{ij}=0$ the evolution operator has an equilibrium solution with
zero eigenvalue $\lambda_0=0$ and
\begin{eqnarray}
    0&=&\sum_{pq}L^{ijpq}V^{pq}_0,\\
    0&=&\sum_{pq}L^{pqij}U^{pq}_0,
\end{eqnarray}
and the inverse is singular. The real part of the second-largest eigenvalue $\Re(\lambda_1)=-\hbar/\tau_{rlx}$ determines the rate of the relaxation
to equilibrium. 
If the system is open $\Gamma^{ij}\neq0$, the largest eigenvalue $\lambda_0=-\hbar/\tau_{esc}$ determines the rate of the escape
of electrons leaking out of the system. If we regard the couplings $\Gamma^{ij}$ as perturbations
of the equilibrium, we can calculate the largest eigenvalue in the first order of perturbation theory as the expectation value
of the perturbation in equilibrium
\begin{equation}
    \lambda_0=-\frac{1}{2}\sum_{ijpq}U^{ij}_0(\Gamma^{ip}\delta_{jq}+\delta_{ip}\Gamma^{qj})V^{pq}_0, \label{lz}
\end{equation}
where $U^{ij}_0$ and $V^{pq}_0$ are the left and right eigenfunctions of the operator
\begin{eqnarray}
    L^{ijpq}_0&=&-i(E_i-E_j)\delta_{ip}\delta_{jq}  
    +\frac{1}{2}\left(\tilde{\gamma}^{ip}+\tilde{\gamma}^{jp}\right)\delta_{ij}\delta_{pq} -\frac{1}{2}\sum_r(\tilde{\gamma}^{ri}+\tilde{\gamma}^{rj})\delta_{ip}\delta_{jq}.
\end{eqnarray}
The left eigenfunction is $U^{ij}_0=\delta_{ij}$ due to probability conservation. The right eigenfunction can be determined from the detailed balance condition $V_0^{ii}\tilde{\gamma}^{ji}=\tilde{\gamma}^{ij}V_0^{jj}$ and normalization.  
Due to the detailed balance (\ref{db}) we get
\begin{equation}
\frac{V_0^{ii}}{V_0^{jj}}=\frac{\cosh^2((\mu-E_j)/2kT)}{\cosh^2((\mu-E_i)/2kT)},
\end{equation}
and the normalized eigenvector is
\begin{equation}
    V_0^{ij}=\frac{\delta_{ij}}{\mathcal{N}}\frac{1}{\cosh^2((\mu-E_i)/2kT)},
\end{equation}
where $\mathcal{N}=\sum_k 1/\cosh^2((\mu-E_k)/2kT)$ is the normalization factor. Using these eigenvectors in (\ref{lz}) we get
\begin{equation}
    \lambda_0=-\frac{1}{\mathcal{N}}\sum_k\frac{\Gamma^{kk}}{\cosh^2((\mu-E_k)/2kT)}.
\end{equation}
As long as the relaxation to equilibrium is much faster than the escape time $\tau_{esc}\gg \tau_{rlx}$, we can use the perturbative first eigenvalue and can neglect the rest of the eigenvalues in the 
expression of the inverse, and we get
\begin{equation}
   [L^{-1}]^{ijpq}\approx \frac{1}{\lambda_0}V_0^{ij}U_0^{pq}\approx -\frac{1}{\sum_k \Gamma^{kk}/\cosh^2((\mu-E_k)/2kT)}\frac{\delta_{ij}\delta_{pq}}{\cosh^2((\mu-E_i)/2kT)}.\label{pertinverse}
\end{equation}
Substituting this expression into the conductance formula (\ref{cndf}) and using (\ref{S14}) yields the 
conductance 
\begin{equation}
    G=\frac{2e^2}{h}T+\frac{2e^2}{\hbar}\frac{{Z}_L{Z}_R}{{Z}_L+{Z}_R}+\frac{e^2}{h}\left[\frac{{Z}_L}{{Z}_L+{Z}_R}T_{R}+\frac{{Z}_R}{{Z}_L+{Z}_R}T_{L}\right]
\end{equation}    
where
\begin{eqnarray}
    {Z}_{L}&=&\sum_k\frac{\Gamma_{L}|\Psi_{L}^k|^2}{4kT\cosh^2((\mu-E_k)/2kT)},\\
    {Z}_{R}&=&\sum_k\frac{\Gamma_{R}|\Psi_{R}^k|^2}{4kT\cosh^2((\mu-E_k)/2kT)},
\end{eqnarray}
and
\begin{eqnarray}
    T_{L}&=&\sum_k \frac{\Gamma_L^2|\Psi_{L}^k|^4}{(\mu-E_k)^2+{(\Gamma^k/2)^2}},\\
    T_{R}&=&\sum_k \frac{\Gamma_R^2|\Psi_{R}^k|^4}{(\mu-E_k)^2+{(\Gamma^k/2)^2}},\\
    T&=&\sum_k \frac{\Gamma_L\Gamma_R|\Psi_{L}^k|^2|\Psi_{R}^k|^2}{(\mu-E_k)^2+{(\Gamma^k/2)^2}},\label{BW}
\end{eqnarray}
where $\Gamma^k=\Gamma_L|\Psi_{L}^k|^2+\Gamma_R|\Psi_{R}^k|^2$. 

\newpage
\section*{S4. Code workflow}
Our starting point is a 3D structure from the Protein Data Bank (\href{https://www.rcsb.org/}{https://www.rcsb.org/}), which we process with the Maestro software (https://www.schrodinger.com/platform/products/maestro/). We connected the appropriate cysteines to the hemes and conducted a force-field minimization procedure only on the side chains of those cysteines to optimize the spatial arrangement of the atoms. Then, C-terminal oxygen atoms and missing hydrogen atoms were added to the protein structures. The outputs of this software are the XYZ files, which contain all the atoms and their coordinates in 3D. Then, the input files (YAH files) suitable for the Yaehmop software (https://yaehmop.sourceforge.net/) are created with a few lines of Python code.
Then, the quantum chemistry calculations were carried out by Yaehmop, using the Extended Hückel method. The outputs are the Hamiltonian and Overlap matrices of the structures in the non-orthogonal Slater-type atomic orbital basis. 
\subsection*{A. Conductance calculation}
First, we transform the Hamiltonian into an orthogonal basis with Löwdin transformation using \\ \texttt{scipy.linalg.fractional\_matrix\_power} in Python. Then, with \texttt{scipy.linalg.eigh}, we solve the eigenvalue problem and get the spectrum and the eigenvectors (molecular orbitals) of the proteins. These two steps are the most resource-consuming ones as they scale with $N^3$, where $N$ is the number of atomic orbitals of the protein structure. After that, we choose the atomic orbitals that are connected to the left and right electrodes (L, R). Utilizing Eqs.(7-9), we lastly calculate the conductance between these selected atomic orbitals with our own Python code.

\subsection*{B. Visualization}

In the original basis, we solve the generalized eigenvalue problem with the Hamiltonian and Overlap matrices in Python with \texttt{scipy.linalg.eigh}. We also made a 3D rectangular grid with $1.5\,$\r{A} resolution with the griData module of the MDAnalysis Python package (https://www.mdanalysis.org/GridDataFormats/). Using the spectrum and the eigenvectors (molecular orbitals), we are able to calculate the functions $\mathcal{T}(\mathbf{r})$ and $\mathcal{Z}(\mathbf{r})$ on the grid. With the griData module, we generate DX files that are suitable outputs for the Visual Molecular Dynamics software (https://www.ks.uiuc.edu/Research/vmd/) to visualize volumetric data and the protein structure itself in 3D.
The conductance calculation and visualization code workflow are summarized in the following chart.

\begin{figure}[h]
    \centering
    \includegraphics[width=0.5\linewidth]{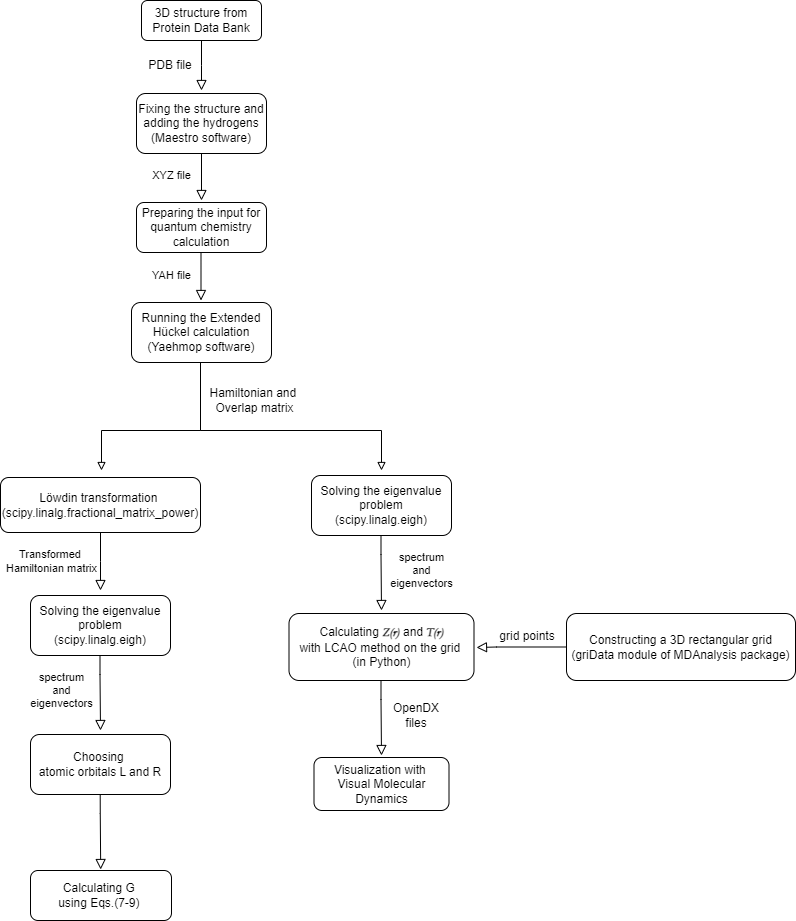}
    \caption{Code workflow for conductance calculation (left branch) and visualization (right branch).}
    \label{fig:enter-label}
\end{figure}

\end{document}